\documentclass[aps,prb,twocolumn]{revtex4-1}

\usepackage{graphicx}
\usepackage{dcolumn}
\usepackage{bm}
\usepackage{booktabs,array,mathptmx,float,tabularx,lipsum,amsmath,multirow}
\usepackage{siunitx,xcolor}
\usepackage{color}
\usepackage{float}
\usepackage{subfig}
\usepackage{epstopdf}
\usepackage{booktabs}
\usepackage{braket}
\usepackage{soul}

\begin{document}


\title{Frequency renormalization and its effects in nonlinear phononics with  $Q_RQ_{IR}^{2}$-type coupling}


\author{Yijie Zeng}
\email[]{zengyj@hdu.edu.cn}
\affiliation{College of Science, Hangzhou Dianzi University, Hangzhou 310018, China}


\date{\today}

\begin{abstract}
A two-phonon system with lowest-order coupling of form $Q_RQ_{IR}^2$ is studied by perturbation method, and analytic results for both phonon displacements and frequencies are obtained. The frequency renormalization of infrared (IR) active mode brings the rectification of Raman mode to saturate at high pump field. For degenerate IR mode with coupling of form $Q_R(Q_{IR,x}^2-Q_{IR,y}^2)$, the frequency of IR mode will split when resonantly pumped by elliptically or linearly polarized ultrashort mid-IR pulse, realizing Raman rectification and magnetization simultaneously. Our results reveal a dynamical effect of nonlinear phononics not captured by first-principles calculation, extend the dynamical multiferroicity to systems with coupling $Q_R(Q_{IR,x}^2-Q_{IR,y}^2)$, and the method can be readily applied to higher-order couplings. The amplitude saturation under strong pump field stimulates future researches to overcome this nonlinear effect.
\end{abstract}


\maketitle


\section{Introduction}
In recent years, due to the development of high-intensity infrared (IR) laser pulses\cite{Liu_B2017,Narrowband2020,pavicevic2025tunablenarrowbandthzgeneration} and the computational efforts\cite{Juraschek2017a,Libbi,levchuk2025nonlinearphononicsbi2te3nanoscale,Dissip} in revealing the anharmonic coupling between IR and other modes (e.g. Raman phonon,  ferroelectric soft phonon mode\cite{Subedi2015,Subedi2017,Ferroelectricity2022}), nonlinear phonon coupling\cite{Disa_2021} was found to exist in a wide variety of materials, and plays an important role in light-induced superconductivity\cite{SC_YBCO_2014}, light-induced magnetism (CoF$_2$)\cite{CoF2_2020}, ultrafast switching of ferroelectricity\cite{LiNbO3_FE_QIR2,SrTiO3_Fechner}, phonon-induced geometric chirality\cite{doi:10.1021/acsnano.4c05978,doi:10.1126/science.adr4713} and multiferroic polarization\cite{7lm1-wm3y}, etc. Compared with other static manipulation methods such as high-pressure, strain or heterostructure engineering, nonlinear phonon coupling is mode-selective, fast (within picoseconds), and often drives the system to a symmetry-breaking state which has no equilibrium counterpart\cite{PhysRevLett.132.016603}.

The physics behind nonlinear phononics is a nonvanishing cubic term $Q_RQ_{IR}^2$ in lattice potential, which couples IR active mode and Raman active (or silent) mode. Here $Q_{R/IR}$ denote the normal coordinates of the coupled Raman active (or silent) and IR active mode, respectively. The irreducible representations of the coupled $Q_R$ and $Q_{IR}$ can be determined by requiring that $\Gamma_R\otimes\Gamma_{IR}^2$ contains the totally symmetric irreducible representation\cite{Radaelli}. When the IR active mode is resonantly pumped by laser pulse of form $F_0e^{-t^2/\sigma^2}cos(\omega_{IR}^{(0)}t)$, the Raman mode oscillates in $2\omega_{IR}$ around a dispaced nonequilibrium position\cite{Subedi_2014}, realizing rectification of Raman mode. Experimentally, the magnitude of rectification is found to scale with square of the electric field of pumped light \cite{RN1360,CoF2_2020,doi:10.1126/science.adr4713}, while it's not clear whether this behavior will survive under stronger laser pulses.  Some theorectical works on HgTe\cite{PhysRevLett.132.016603}, Mott insulating titanites YTiO$_3$\cite{PhysRevB.98.024102} and YBa$_2$Cu$_3$O$_7$\cite{PhysRevB.94.134307} give hints that the rectification tends to saturate at high pump field, yet whether it is a general feature of cubic anharmonicity cannot be determined affirmatively. These facts arouse the interesting question --- is it possible to drive the Raman mode to any desired rectified position with stronger pump pulse?

It is well-known that in one-phonon system with anharmonic cubic term, the frequency is a function of amplitude\cite{PhysRevLett.88.067401,RN1874}. The cubic term in two-phonon system should also cause the frequencies of both phonons to shift with amplitudes, leading to ``frequency renormalization''. Although there were reports about frequency renormalization of IR active mode\cite{PhysRevB.94.134307,PhysRevMaterials2064401}, its influence on the dynamics of the two-phonon system, especially the rectification of Raman mode is yet to be uncovered. Here we tackle this problem by deriving an analytical expression of frequencies of IR and Raman modes ($\omega_{IR}$ and $\omega_{R}$) as a function of their eigen amplitudes ($a_R$ and $a_{IR}$), and analyzing its effect on Raman rectification in detail. As IR mode red-shifts under strong pump field, the linear increase of $a_{IR}$  with pump field amplitude $E_0$ will be suppressed, leading to a saturation in the rectification amplitude of Raman active phonon. In Sec.\ref{Results_A} the frequency-amplitude relation is derived, following the perturbation method suggested by Landau and Lifshitz\cite{book1}. This method is analytically solvable, cheap and materials independent, compared with other realistic method, \textit{e.g. ab initio} molecular dynamics\cite{PhysRevLett.132.016603}. In Sec.\ref{Results_B} we discuss the $a_{IR}$-$E_0$ relation, numerical results of Raman rectification as a function of $E_0$ are given, along with some new features not revealed by analytical method. When the IR mode is degenerate, as shown in Sec.\ref{Multiferroicity}, the frequency of IR mode splits when resonantly pumped by elliptically or linearly polarized ultrashort mid-IR laser pulse, and Raman rectification and (oscillating) magnetization can be induced simultaneously. Finally, the frequency-amplitude relation can also be deduced from quantum perturbation method. 

Nonlinear phononics makes use of tailored laser fields to resonantly drive the IR active phonons to large amplitudes\cite{nonlinear_phononics_2022}. Our finding that the resonantly driven IR amplitude (and the induced rectification of Raman active phonon) tends to saturate under strong pump field indicates a dynamical constraint for accessible phonon displacement in quantum materials via nonlinear phononics. The amplitude saturation can be extended to system with strong spin-lattice interaction\cite{FePS3_A,lu2025ultrafastcooperativeelectronicstructural}, and is more easily observed in systems with large coupling constant, e.g. the coupling between interlayer sliding phonon and intralayer shear phonon in WTe$_2$\cite{horstmann2025coherentphononcontrolamplitude}.  One way to improve the saturation amplitude is to use off-resonant laser field with slightly reduced frequency. The laser-induced dynamical multiferroicity\cite{7lm1-wm3y,2sj3-33ky} is expected to occur in systems with phonon nonlinearity both in form $Q_R(Q_{IR,x}^2-Q_{IR,y}^2)$ and $Q_RQ_{IR,x}Q_{IR,y}$, extending the candidate materials for future ultrafast data processing technology.

\section{Results and Discussion}

\subsection{The frequency-amplitude relation with $Q_RQ_{IR}^{2}$ coupling}
\label{Results_A}
Assume the IR active phonon is nondegenerate and has no associated phonon angular momentum\cite{chiral_phonon_1,chiral_phonon_2}. In the absence of damping, the Lagrangian of the two phonon system can be written as
\begin{equation}
L=\frac{1}{2}\sum_{i}\dot{Q}_i^{2}-\frac{1}{2}\sum_{i}\omega_{i}^{(0)2}Q_{i}^2+\frac{1}{2}gQ_{R}Q_{IR}^2+Q_{IR}F(t)
\end{equation}
where $i$ runs over $\{R, IR\}$, the third term on the right denotes the coupling between Raman and IR modes, and the fourth term is the potential energy caused by laser pulse. To emphasize the renormalization of frequency, we use $\omega_{i}^{(0)}$ to denote the natural angular frequency, and $\omega_i$ to denote the observed angular frequency of the $i$ mode. We assume $F(t)=Z_e^{*}E_0e^{-t^2/\sigma^2}cos(\gamma t)$, which corresponds to a pump pulse with Gaussian distribution. Here $Z_e^{*}$ is Born effective charge of the IR mode, $E_0$, $\sigma$ and $\gamma$ are the largest magnitude of the electric field, the time-domain duration, and the angular frequency of the laser pulse, respectively. Lagrangian's equation on $Q_R$ and $Q_{IR}$ then give

\begin{subequations}
	\label{Eq_EOM}
	\begin{equation}
	\label{Eq_EOM_a}
	\ddot{Q}_R+\omega_{R}^{(0)2}Q_R=\frac{1}{2}gQ_{IR}^2
	\end{equation}
	\begin{equation}
	\label{Eq_EOM_b}
	\ddot{Q}_{IR}+\omega_{IR}^{(0)2}Q_{IR}=gQ_RQ_{IR}+F(t)
	\end{equation}
\end{subequations}

\begin{figure}[h]
	\includegraphics[width=6cm]{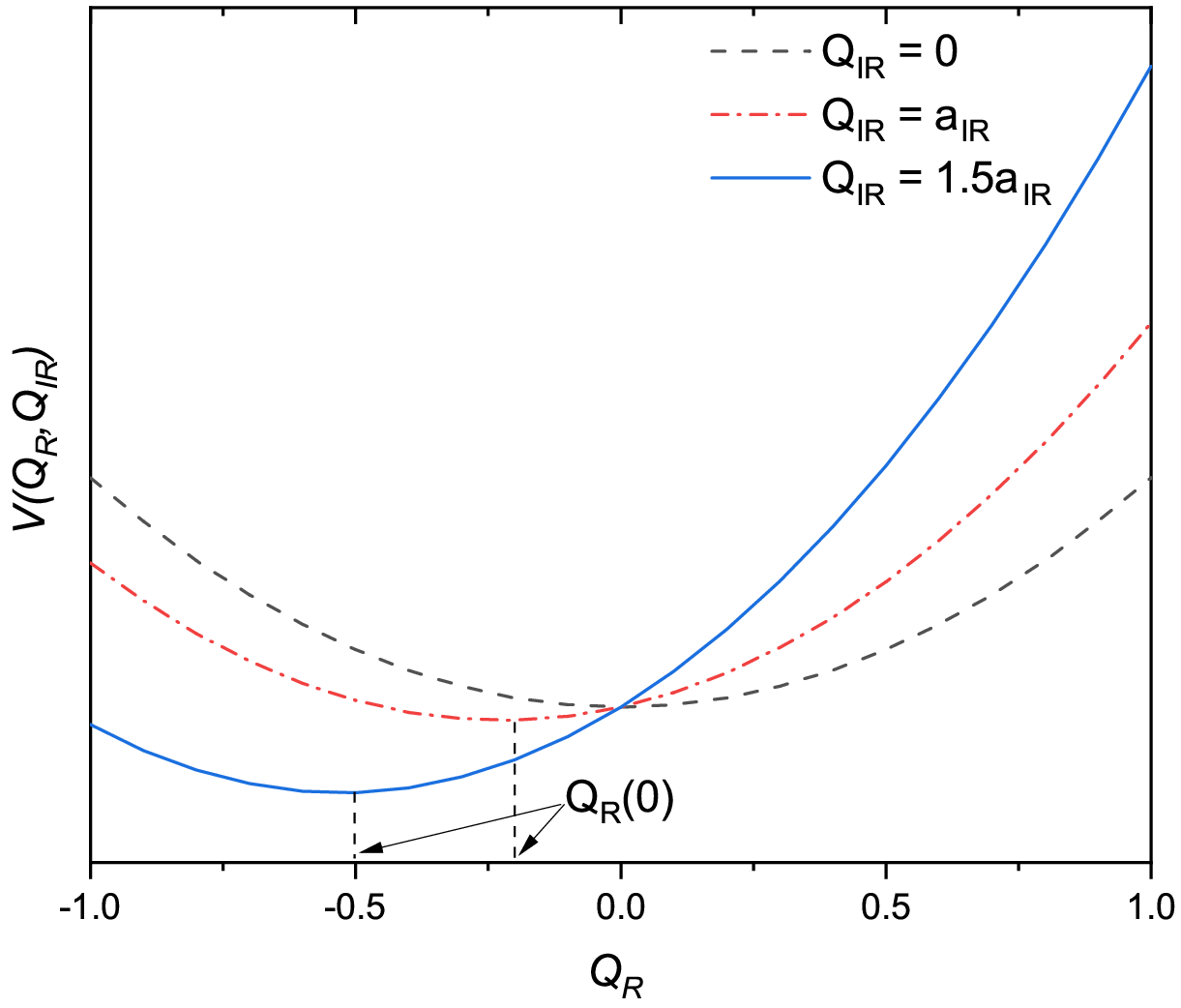}
	\caption{The lattice potential of a two-phonon system with cubic term $Q_RQ_{IR}^2$. The rectification of Raman mode $Q_R(0)$ increases with the displacement of IR mode.}
	\label{Fig_illustration}
\end{figure}

As a first step, we ignore $F(t)$ in Eq.(\ref{Eq_EOM_b}), namely we seek the solution at $t\rightarrow\infty$, \textit{i.e.} the steady-state solution. Following Landau and Lifshitz\cite{book1}, we obtain the following frequency-amplitude relations, which is valid to the second-order of coupling constant $g$, 

\begin{subequations}
		\label{Eq_Sol_shift_freq}
		\begin{equation}
			\omega_{R}=\omega_{R}^{(0)}-\frac{g^2a_{IR}^2}{4\omega_{R}^{(0)}}\frac{1}{(2\omega_{IR}^{(0)})^2-(\omega_{R}^{(0)})^2}
		\end{equation}
		\begin{equation}
		\begin{aligned}
			\omega_{IR}&=\omega_{IR}^{(0)}-\frac{g^2a_{IR}^2}{8\omega_{IR}^{(0)}}[\frac{1}{\omega_{R}^{(0)2}}-\frac{1}{2}\frac{1}{(2\omega_{IR}^{(0)})^2-(\omega_{R}^{(0)})^2}] \\
			&-\frac{g^2a_{R}^2}{4\omega_{IR}^{(0)}}\frac{1}{(2\omega_{IR}^{(0)})^2-(\omega_{R}^{(0)})^2}
		\end{aligned}
		\end{equation}
\end{subequations}
where $a_{IR/R}$ denote the zero-order oscillating amplitude of IR/Raman active mode.  When condition $\sqrt{2}\omega_{IR}^{(0)}\gg \omega_{R}^{(0)}$ is satisfied, which is the usual case, the angular frequencies will decrease as the amplitudes increase. Eq.(\ref{Eq_Sol_shift_freq}) allows us to infer the oscillation amplitudes $a_{IR}$ and $a_{R}$ from the frequencies $\omega_{R}$ and $\omega_{IR}$ if $g$ is known, and is one of the main results of this work. We note that Fechner \textit{et. al.}\cite{PhysRevMaterials2064401} had obtained a similar result (see Eq.(A1) and (A2) in the cited reference), which differs from ours in several places: (1) They considered a sinusoidal driving force, while a pulsed force is considered here. (2) The second term on the right of their Eq.(A1) is proportional to $a_R^2$, while ours is $a_{IR}^2$, and their fourth term is proportional to $a_{IR}$, which should be $a_{IR}^2$. (3) The third term on the right of their Eq.(A2) is proportional to $1/\omega_{IR}$, which should be $1/\omega_{IR}^2$ by dimension analysis. To second-order, the IR mode will oscillate in a superposition state with frequencies $\omega_{IR}$, $\omega_{IR}\pm\omega_{R}$, $\omega_{IR}\pm2\omega_{R}$ and $3\omega_{IR}$, while the Raman mode oscillates with frequencies  $\omega_{R}$, $0$, $2\omega_{IR}$ and $2\omega_{IR}\pm\omega_{R}$, where $0$ is the rectified component, as shown schematically in Fig.\ref{Fig_illustration}. The amplitudes of each component are listed in Table.\ref{tab:Frequency_Amplitude}. The detailed deduction can be found in Appendix.\ref{Apend1}. 

\begin{table}[H]
	\caption{The angular frequencies and amplitudes of oscillation components for $Q_R$ and $Q_{IR}$, to the second-order of coupling constant $g$. The expressions for $\omega_{R}$ and $\omega_{IR}$ are given by Eqs.(\ref{Eq_Sol_shift_freq}).}
	\begin{ruledtabular}
		\begin{tabular}{ccc}
			Term & Angular frequency  & Amplitude  \\
			\hline
			$Q_R^{(0)}$ & $\omega_{R}$ & $a_R$  \\
			$Q_R^{(1)}$ & $0$ & $\frac{\frac{1}{4}ga_{IR}^2}{\omega_{R}^{(0)2}}$  \\
			 & $2\omega_{IR}$ & $\frac{\frac{1}{4}ga_{IR}^2}{\omega_{R}^{(0)2}-(2\omega_{IR}^{(0)})^2}$  \\
			 $Q_R^{(2)}$ & $2\omega_{IR}-\omega_{R}$ & $\frac{\frac{1}{4}g^2a_Ra_{IR}^2}{[\omega_{IR}^{(0)2}-(\omega_{IR}^{(0)}-\omega_{R}^{(0)})^2][\omega_{R}^{(0)2}-(2\omega_{IR}^{(0)}-\omega_{R}^{(0)})^2]}$  \\
			  & $2\omega_{IR}+\omega_{R}$ & $\frac{\frac{1}{4}g^2a_Ra_{IR}^2}{[\omega_{IR}^{(0)2}-(\omega_{IR}^{(0)}+\omega_{R}^{(0)})^2][\omega_{R}^{(0)2}-(2\omega_{IR}^{(0)}+\omega_{R}^{(0)})^2]}$  \\
			  \dots & \dots & \dots \\
			  \hline
			  $Q_{IR}^{(0)}$ & $\omega_{IR}$ & $a_{IR}$  \\
			  $Q_{IR}^{(1)}$ & $\omega_{IR}-\omega_{R}$ & $\frac{\frac{1}{2}ga_Ra_{IR}}{\omega_{IR}^{(0)2}-(\omega_{IR}^{(0)}-\omega_{R}^{(0)})^2}$  \\
			   & $\omega_{IR}+\omega_{R}$ & $\frac{\frac{1}{2}ga_Ra_{IR}}{\omega_{IR}^{(0)2}-(\omega_{IR}^{(0)}+\omega_{R}^{(0)})^2}$  \\
			   $Q_{IR}^{(2)}$ & $\omega_{IR}-2\omega_{R}$ & $\frac{\frac{1}{4}g^2a_R^2a_{IR}}{[\omega_{IR}^{(0)2}-(\omega_{IR}^{(0)}-\omega_{R}^{(0)})^2][\omega_{IR}^{(0)2}-(\omega_{IR}^{(0)}-2\omega_{R}^{(0)})^2]}$  \\
			    & $\omega_{IR}+2\omega_{R}$ & $\frac{\frac{1}{4}g^2a_R^2a_{IR}}{[\omega_{IR}^{(0)2}-(\omega_{IR}^{(0)}+\omega_{R}^{(0)})^2][\omega_{IR}^{(0)2}-(\omega_{IR}^{(0)}+2\omega_{R}^{(0)})^2]}$  \\
			     & $3\omega_{IR}$ & $\frac{\frac{1}{8}g^2a_{IR}^3}{[\omega_{R}^{(0)2}-(2\omega_{IR}^{(0)})^2][\omega_{IR}^{(0)2}-(3\omega_{IR}^{(0)})^2]}$  \\
			     \dots & \dots & \dots \\
		\end{tabular}
	\end{ruledtabular}
	\label{tab:Frequency_Amplitude}
\end{table}

\subsection{The variation of amplitude of eigen IR mode with pump field}
\label{Results_B}
\subsubsection{Linear increase of $a_{IR}$ with $E_0$ under small field amplitude}
In the previous section we've seen that the zero-order amplitudes $a_R$ and $a_{IR}$ determine all higher-order amplitudes, it is thus necessary to determine how $a_R$ and $a_{IR}$ vary with the amplitude of pump field $E_0$. In other words, we wish to determine $a_R$ and $a_{IR}$ under different $E_0$ in Eq.(\ref{Eq_EOM}). Suppose initially both $a_R$ and $a_{IR}$ equals to zero, and $a_{IR}$ increases linearly with the pump field amplitude $E_0$, the validity of which will be soon verified, then $Q_{IR}$ is proportional to $E_0$ and  $Q_{R}$ is proportional to $E_0^2$, leading to $gQ_RQ_{IR}$ proportional to $E_0^3$. As $F(t)$ is proportional to $E_0$, under small pump field amplitude $F(t)$ is much larger than $gQ_RQ_{IR}$. The latter can safely be neglected and Eq.(\ref{Eq_EOM}) becomes

\begin{subequations}
	\label{Eq_EOM_b1}
	\begin{equation}
	\label{Eq_EOM_a_1}
	\ddot{Q}_R+\omega_{R}^{(0)2}Q_R=\frac{1}{2}gQ_{IR}^2
	\end{equation}
	\begin{equation}
	\label{Eq_EOM_b_1}
	\ddot{Q}_{IR}+\omega_{IR}^{(0)2}Q_{IR}=Z_e^{*}E_0e^{-t^2/\sigma^2}cos(\gamma t)
	\end{equation}
\end{subequations}

It's then obvious that under small pump field, the IR mode will oscillate with frequency $\omega_{IR}^{(0)}$, while the Raman mode will be in first-order state $Q_R^{(1)}$, or oscillate around a displaced position with frequency $2\omega_{IR}^{(0)}$. In the following we show the oscillation amplitude $a_{IR}$ of IR mode is proportional to $E_0$.

When $t$ is located in $[-\sigma,+\sigma]$, the external force is largest and the oscillating frequency of $Q_{IR}(t)$ is $\gamma/(2\pi)$ --- the frequency of the laser pulse. As $t$ increases, the force diminishes quickly, and $Q_{IR}$ restores to oscillate with its normal frequency $\omega_{IR}^{(0)}$. The long-time oscillation amplitude can be obtained from energy conservation\cite{book1}: The energy pumped in by the external force is half the squared modulus of Fourier component $\frac{1}{2}|F(\omega)|^2$  of the force at $\omega_{IR}^{(0)}$, where $F(\omega)$ is the Fourier transform of the force

\begin{equation}
\label{Eq_FT_Force}
\begin{aligned}
F(\omega)&=\int_{-\infty}^{+\infty}Z_e^{*}E_0e^{-t^2/\sigma^2}cos(\gamma t)e^{-i\omega t}dt \\
&=\frac{\sqrt{\pi}}{2}Z_e^{*}E_0\sigma[e^{-\frac{\sigma^2}{4}(\gamma-\omega)^2}+e^{-\frac{\sigma^2}{4}(\gamma+\omega)^2}]
\end{aligned}
\end{equation}

The energy pumped in should equal to the oscillation energy $\frac{1}{2}\omega_{IR}^{(0)2}a_{IR}^2$. For typical values of $\sigma$ ($\sim 0.1$ ps) and $\gamma$ ($\sim 2\pi\times 10$ THz, same for $\omega$), the second term ($\sim e^{-4\pi^2}$) in Eq.(\ref{Eq_FT_Force}) can be neglected. Thus we have 

\begin{equation}
    \label{Eq_aIR_without_g}
	a_{IR}(\gamma)=\frac{\sqrt{\pi}Z_e^{*}E_0\sigma}{2\omega_{IR}^{(0)}} e^{-\frac{\sigma^2}{4}(\gamma-\omega_{IR}^{(0)})^2}
\end{equation}

The linear relation between $a_{IR}$ and $E_0$ is evident. In previous works\cite{Subedi_2014,Radaelli}, $\gamma=\omega_{IR}^{(0)}$ were assumed. Eq.(\ref{Eq_aIR_without_g}) shows that the oscillation amplitude is largest when $\gamma=\omega_{IR}^{(0)}$ and decreases as $\gamma$ deviates from $\omega_{IR}^{(0)}$. The full width at half maximum (FWHM) of $F(\omega)$ is $4\sqrt{ln2}/\sigma$. Here the duration of the pulse $\sigma$ should be chosen carefully: A small $\sigma$ is required to obtain large amplitude in a wider frequency region, but at the same time it reduces the amplitude.

\subsubsection{Saturation of $a_{IR}$ with $E_0$ under large field amplitude}

\begin{figure}[h]
	\includegraphics[width=8cm]{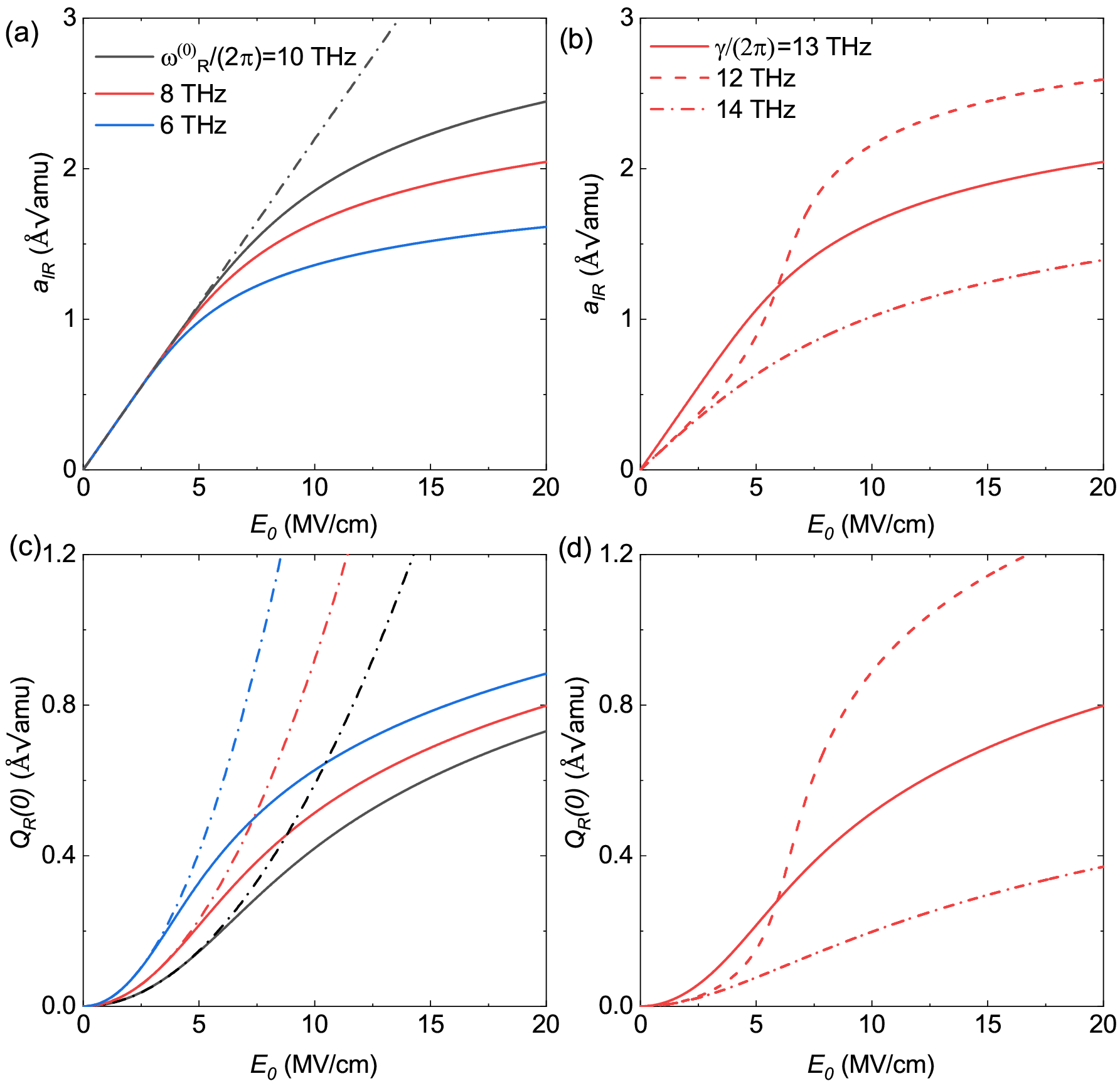}
	\caption{The amplitudes of (a,b) eigen IR mode and (c,d) the corresponding rectified Raman mode as a function of pulse amplitude. $\omega_{IR}^{(0)}/(2\pi)=\gamma/(2\pi)=13$ THz in (a) and (c), and $\omega_{R}^{(0)}/(2\pi)=8$ THz, $\omega_{IR}^{(0)}/(2\pi)=13$ THz in (b) and (d).  The dashed lines in (a) and (c) are linear and quadratic fitting curves, respectively.} 
	\label{Fig_aIR_E0}
\end{figure}

As the pump field amplitude increases, the $gQ_RQ_{IR}$ term increases more quickly than $F(t)$ and cannot be ignored. The inclusion of this term can be regarded as a renormalization of frequency\cite{RN1356} of the IR mode ($\omega_{IR}^{(0)}\rightarrow\sqrt{\omega_{IR}^{(0)2}-gQ_R}$), as long as the eigen Raman mode does not appear. Thus Eq.(\ref{Eq_EOM_b1}) can still be applied, but with $\omega_{IR}^{(0)}$ replaced by $\omega_{IR}$, as given by Eq.(\ref{Eq_Sol_shift_freq}b). For $\gamma=\omega_{IR}^{(0)}$, the frequency renormalization has the effect of decreasing the proportionality coefficient between $a_{IR}$ and $E_0$. By inserting Eq.(\ref{Eq_Sol_shift_freq}b) (with $a_R=0$) into Eq.(\ref{Eq_aIR_without_g}) we get a transcendental equation about $a_{IR}$

\begin{equation}
    \label{Eq_aIR_with_g}
	a_{IR}(1-\frac{c}{\omega_{IR}^{(0)}}a_{IR}^2)=\frac{\sqrt{\pi}Z_e^{*}E_0\sigma}{2\omega_{IR}^{(0)}} e^{-\frac{\sigma^2}{4}(\gamma-\omega_{IR}^{(0)}+ca_{IR}^2)^2}
\end{equation}
where Eq.(\ref{Eq_Sol_shift_freq}b) is rewritten as $\omega_{IR}=\omega_{IR}^{(0)}-ca_{IR}^{2}$ and $c=\frac{g^2}{8\omega_{IR}^{(0)}}[\frac{1}{\omega_{R}^{(0)2}}-\frac{1}{2}\frac{1}{(2\omega_{IR}^{(0)})^2-(\omega_{R}^{(0)})^2}]$ is a constant for given $\omega_{IR}^{(0)}$, $\omega_{R}^{(0)}$ and $g$. Eq.(\ref{Eq_aIR_with_g}) can be solved numerically, and typical result is shown in Fig.\ref{Fig_aIR_E0}(a), where the trend of saturation of $a_{IR}$ as $E_0$ increases is evident. Then by consulting Table.\ref{tab:Frequency_Amplitude}, the rectification of Raman mode  $Q_{R}(0)$  will deviate from quadratic behavior as $E_0$ increases, as shown in Fig.\ref{Fig_aIR_E0}(c). It is possible to realize larger rectification by using a pulse which is slightly off-resonant with the IR mode, at the cost of reducing $a_{IR}$ at small $E_0$, see Fig.\ref{Fig_aIR_E0}(b,d). This behavior was found in YTiO$_3$\cite{PhysRevB.98.024102} previously, which can be extended to all two-phonon system with $Q_RQ_{IR}^{2}$ coupling and suitable $\omega_{IR}^{(0)}$, $\omega_{R}^{(0)}$.

\subsubsection{The numerical results}
\label{Results_C}
\begin{widetext}	

	\begin{figure}[h]
	       \centering
		\includegraphics[width=14cm]{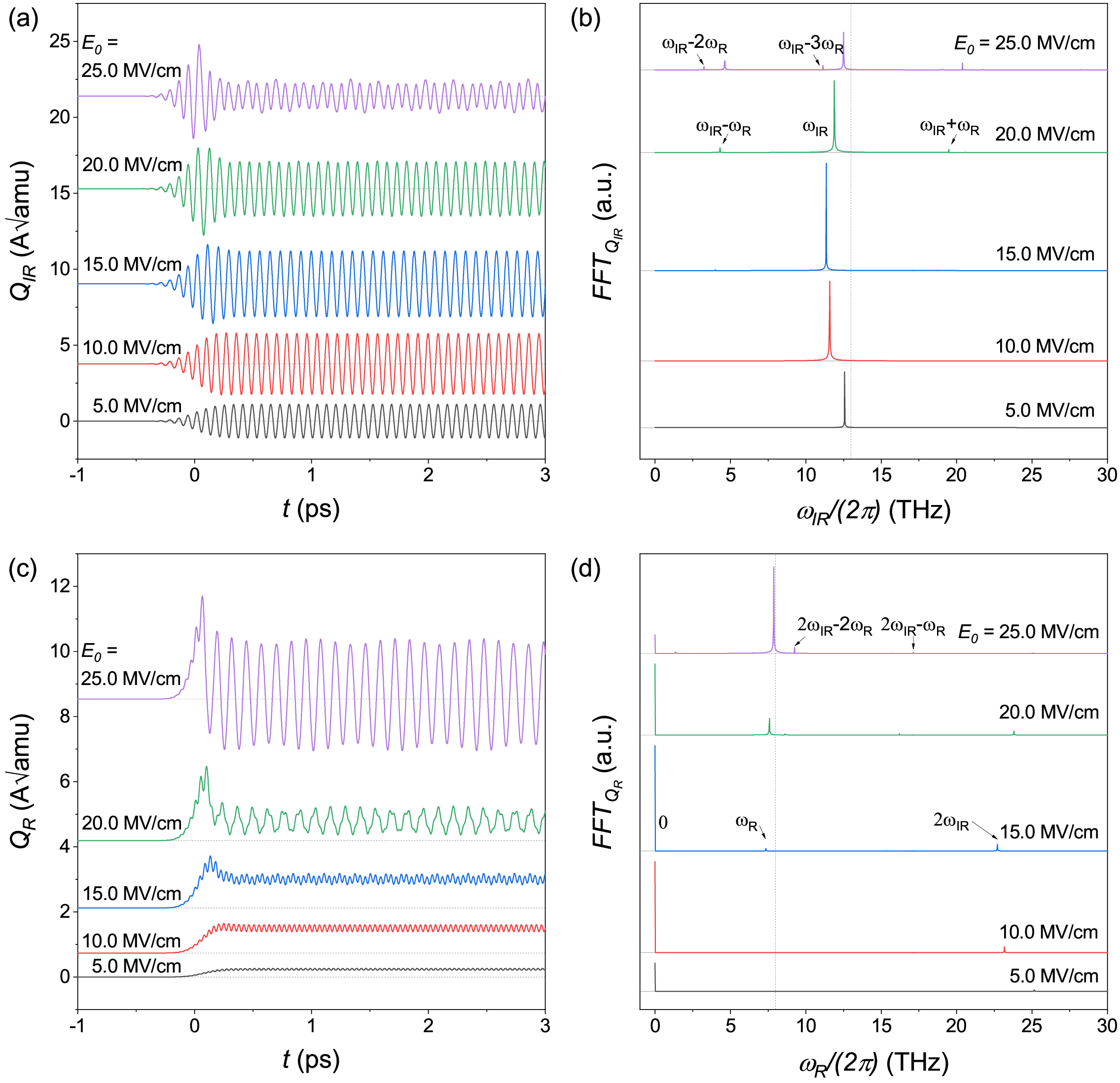}
		\caption{ The transient amplitudes of (a) IR and (c) Raman modes under pump pulses with different amplitudes, where damping effect is ignored. (b) and (d) are the amplitudes of the discrete Fourier transform of (a) and (c), respectively. The dashed lines in (b) and (d) indicate the natural frequencies of IR ($13.0$ THz) and Raman ($8.0$ THz) modes.}
		\label{Fig_QR_QIR_t_without_damping}
	\end{figure}
\end{widetext}

To verify the above picture, we carry out numerical solutions of Eq.(\ref{Eq_EOM}) with initial condition $(Q_{R},\dot{Q}_{R},Q_{IR},\dot{Q}_{IR})|_{t_0}=(0,0,0,0)$  by fourth-order Runge-Kutta method. Here $t_0$ is a negative time where $F(t)$ tends to zero. The parameters used are $\omega_{IR}^{(0)}/(2\pi)=13$ THz, $\gamma=\omega_{IR}^{(0)}$,  $\omega_{R}^{(0)}/(2\pi)=8$ THz, $g=0.2$ eV/$(\sqrt{amu}\AA)^3$ (a large $g$ is used to illustrate the frequency shift, $amu$ denotes the atomic mass unit), $\sigma=0.21$ ps, $Z_e^{*}=1.0$ e/$\sqrt{amu}$ and $E_0$ varies from $0$ to $35$ MV/cm. The time step is $\Delta t=0.625$ fs and $t_0=-10\times{2\pi}/{\omega_{R}^{(0)}}$. Typical results are shown in Fig.\ref{Fig_QR_QIR_t_without_damping}.

Under small amplitude of pump pulse (\textit{e.g.} $E_0=5$ MV/cm), the IR mode oscillates in eigen mode with frequency $\omega_{IR}$ and Raman mode oscillates in a superposition state of $2\omega_{IR}$ and $0$, which is evident in the Fourier transform of the long-time solution, as shown in Fig.\ref{Fig_QR_QIR_t_without_damping}(b,d). This is consistent with the results in Sec.\ref{Results_A} for $a_R=0$, which indicates that only $Q_{IR}^{(0)}$ and $Q_R^{(1)}$ do not vanish. (The $3\omega_{IR}$ term in $Q_{IR}^{(2)}$ is also nonzero, but its magnitude is too small to be discerned.) As the pump pulse amplitude increases, the frequency of eigen IR mode red-shifts, which is clearly seen in Fig.\ref{Fig_QR_QIR_t_without_damping}(b),  and accordingly the $2\omega_{IR}$-component of Raman mode also red-shifts, as shown in Fig.\ref{Fig_QR_QIR_t_without_damping}(d).  

\begin{figure}[h]
	\includegraphics[width=8cm]{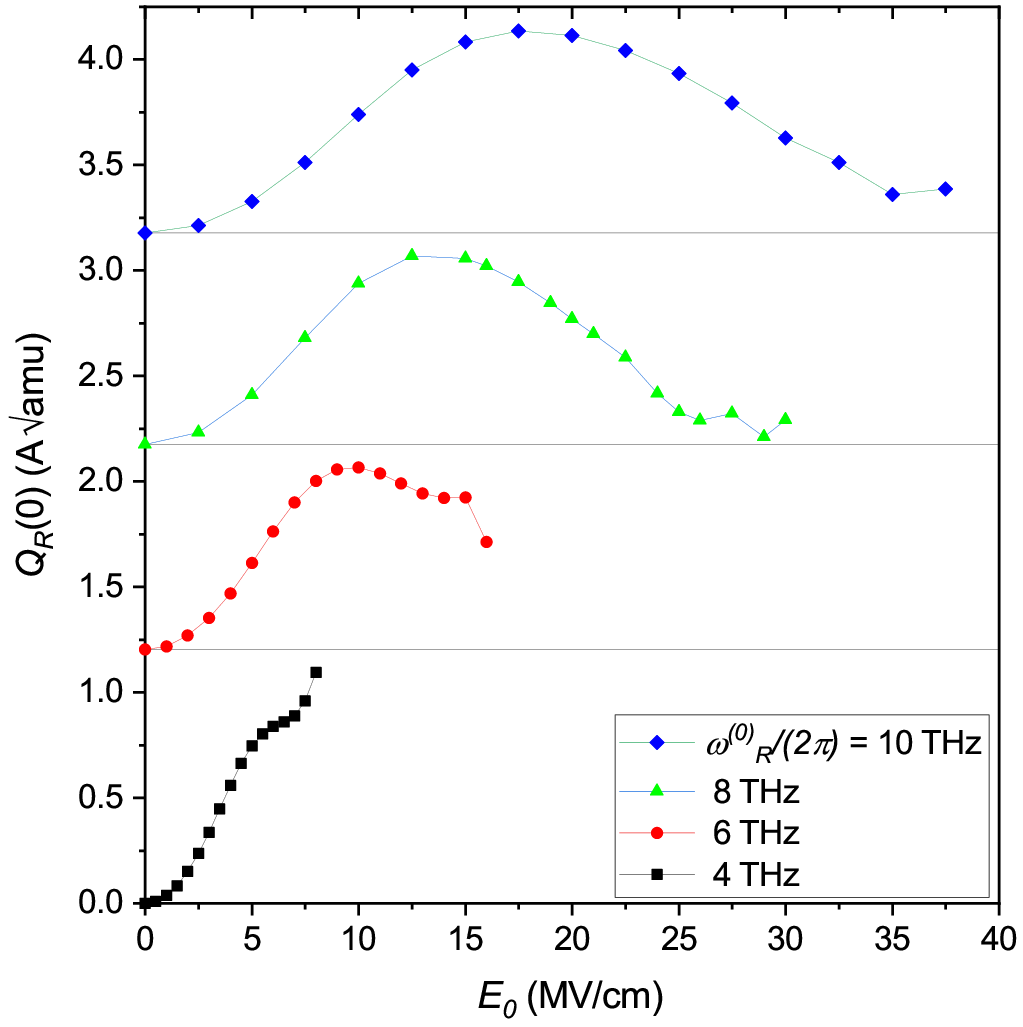}
	\caption{ (a) The variation of rectified Raman mode $Q_R(0)$ with $E_0$ for different frequencies of Raman mode, under resonant pumping. $\omega_{IR}^{(0)}/(2\pi)=13$ THz.}
	\label{Fig_aIR_aR_E0}
\end{figure}

The numerical results reveal two new phenomena: (1) As the amplitude of the pump pulse increases and becomes larger than a critical value (about $15$ MV/cm here), the eigen Raman mode $Q_R^{(0)}$ with frequency $\omega_{R}$ will appear, as is evident in Fig.\ref{Fig_QR_QIR_t_without_damping}(d), and afterwards nearly all the terms up to second-order (even third-order for IR phonon) appear. (2) The appearance of eigen Raman mode causes two possible results, The rectified Raman mode $Q_{R}(0)$ either reaches a maximum and then decreases with $E_0$ (\textit{e.g.} for $\omega_R^{(0)}/(2\pi)=8$ THz), or saturates in intermediate $E_0$ and continues to increase at larger $E_0$ (\textit{e.g.} for $\omega_R^{(0)}/(2\pi)=4$ THz), until the breakup of the system, as shown in Fig.\ref{Fig_aIR_aR_E0}. Which one will happen seems to be determined by $\omega_{R}^{(0)}$ and $\omega_{IR}^{(0)}$, and the required pump field to realize saturation increases with $\omega_{R}^{(0)}$ for a given IR mode. Furthermore, when $\omega_{R}^{(0)}>\omega_{IR}^{(0)}$ and $\omega_{IR}^{(0)}\approx\frac{1}{2}\omega_{R}^{(0)}$, the rectification of Raman mode is weak, and another phenomena --- the ionic Raman scattering occurs, as exemplified in Appendix.\ref{Append_ionicRaman}.

\subsection{Dynamically induced multiferroicity for degenerate IR mode}
\label{Multiferroicity}
We proceed to show that if the IR mode is doubly degenerate and the coupling term is of form $Q_R(Q_{IR,x}^2-Q_{IR,y}^2)$, \textit{e.g.} $B_1$ and $E$ modes in materials with point group $D_{2d}$\cite{Radaelli}, then it's possible to realize Raman rectification and magnetization transiently by resonantly exciting the IR mode with elliptically or linearly polarized mid-IR laser pulse, similar to the mechanism suggested by Paiva \textit{et. al.}\cite{7lm1-wm3y} recently. The pulse-induced multiferroicity discussed below assumes the following conditions to be satisfied: (1) The phonons have coupling of form $Q_R(Q_{IR,x}^2-Q_{IR,y}^2)$. (2) Inversion symmetry is broken in the nonequilibrium state along the displaced Raman mode, and polarization is assumed to scale with the rectification amplitude of the Raman mode. (3) Meanwhile, the driven degenerate IR mode is expected to have large effective magnetic field\cite{phonon_induced_magnetization1}, or large phonon magnetic moment\cite{multiferr_SrTiO3} to make the oscillating magnetization observable. The equation of motion is 

\begin{subequations}
	\label{Eq_EOM_2}
	\begin{equation}
	\label{Eq_EOM_2_a}
	\ddot{Q}_R+\omega_{R}^{(0)2}Q_R=\frac{1}{2}g(Q_{IR,x}^2-Q_{IR,y}^2)
	\end{equation}
	\begin{equation}
	\label{Eq_EOM_2_b}
	\ddot{Q}_{IR,x}+\omega_{IR}^{(0)2}Q_{IR,x}=gQ_RQ_{IR,x}+F_x(t)
	\end{equation}
	\begin{equation}
	\label{Eq_EOM_2_c}
	\ddot{Q}_{IR,y}+\omega_{IR}^{(0)2}Q_{IR,y}=-gQ_RQ_{IR,y}+F_y(t)
	\end{equation}
\end{subequations}

For simplicity, we assume $F_x=eE_x(t)$ and $F_y=eE_y(t)$, where $E_x(t)=E_0cos(\phi)e^{-t^2/\sigma^2}cos(\gamma t)$ and $E_y(t)=E_0sin(\phi)e^{-t^2/\sigma^2}cos(\gamma t+\pi/2)$, with $\gamma=\omega_{IR}^{(0)}$. By the same method outlined in Appendix.\ref{Apend1} we obtain the frequencies and amplitudes of each mode, as summarized in Table.\ref{tab:Frequency_Amplitude_2} in Appendix.\ref{Apend3}. The frequency-amplitude relation of the IR mode is

\begin{subequations}
		\label{Eq_Sol_shift_freq_2}
		\begin{equation}
			\begin{aligned}
			\omega_{IR,x}&=\omega_{IR}^{(0)}-\frac{g^2(a_{x}^2-a_{y}^2)}{8\omega_{IR}^{(0)}\omega_{R}^{(0)2}}+\frac{g^2a_{x}^2}{16\omega_{IR}^{(0)}[(2\omega_{IR}^{(0)})^2-(\omega_{R}^{(0)})^2]} \\
			&-\frac{g^2a_{R}^2}{4\omega_{IR}^{(0)}}\frac{1}{(2\omega_{IR}^{(0)})^2-(\omega_{R}^{(0)})^2}
		\end{aligned}
		\end{equation}
		\begin{equation}
		\begin{aligned}
			\omega_{IR,y}&=\omega_{IR}^{(0)}+\frac{g^2(a_{x}^2-a_{y}^2)}{8\omega_{IR}^{(0)}\omega_{R}^{(0)2}}+\frac{g^2a_{y}^2}{16\omega_{IR}^{(0)}[(2\omega_{IR}^{(0)})^2-(\omega_{R}^{(0)})^2]} \\
			&-\frac{g^2a_{R}^2}{4\omega_{IR}^{(0)}}\frac{1}{(2\omega_{IR}^{(0)})^2-(\omega_{R}^{(0)})^2}
		\end{aligned}
		\end{equation}
	\end{subequations}
where the first terms on the right have opposite signs. The second term is usually much smaller than the first term for $\omega_{IR}^{(0)}>\omega_{R}^{(0)}$, and the third term does not occur at small pump field ($a_{R}=0$). In the circumstance $a_x\neq a_y$, the original degenerate IR mode will become nondegenerate.

\begin{widetext}	

	\begin{figure}[h]
	       \centering
		\includegraphics[width=16cm]{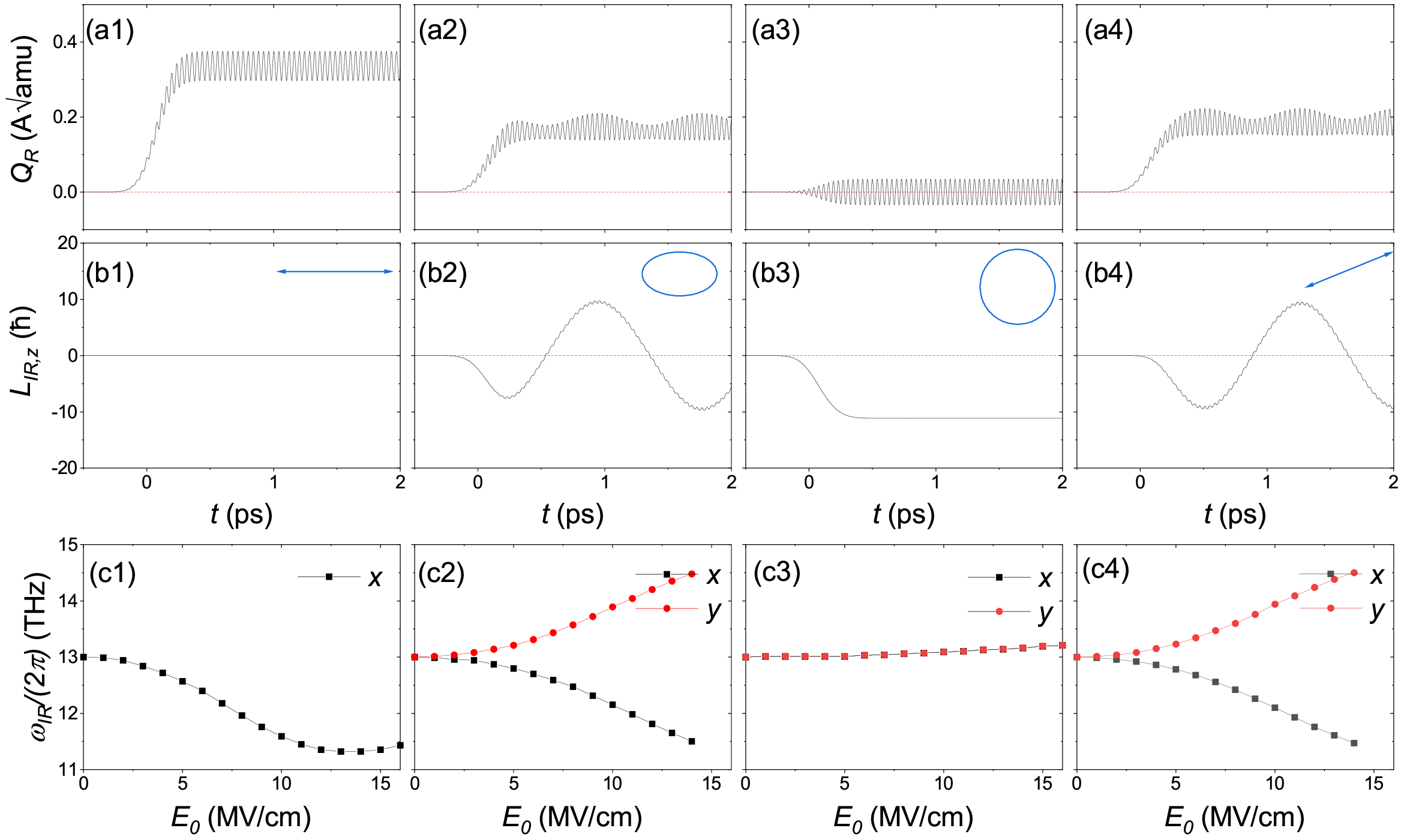}
		\caption{ The transient amplitudes of (a) Raman mode and (b) angular momentum of IR modes under pump pulses with amplitude $E_0=6$ MV/cm and (1) linear polarization along $x$, (2) elliptical polarization with $\phi=\pi/6$, (3) circular polarization and (4) linear polarization $\pi/6$ with respect to $x$. (c) The frequency of IR mode as function of pump pulse amplitude.}
		\label{Fig_QIRx2_y2}
	\end{figure}
\end{widetext}

For linearly polarized mid-IR pulse along $x$ ($\phi=0$) or $y$ ($\phi=\pi/2$) direction, only $Q_{IR,x}$ or $Q_{IR,y}$ will be nonzero, and Eq.(\ref{Eq_EOM_2}) returns to Eq.(\ref{Eq_EOM}). The rectification of Raman mode will be switched between $x$ and $y$ polarized light, while the IR mode has no phonon angular momentum, see Fig.\ref{Fig_QIRx2_y2}(a1,b1,c1).

When pumped by elliptically polarized mid-IR pulse ($0<\phi<\pi/4$), $Q_{IR,x}$ and $Q_{IR,y}$ are both excited, but to different amplitudes ($a_x>a_y$), the Raman mode will be oscillating in $2\omega_{IR,x}$ and $2\omega_{IR,y}$ around a rectified position $\langle Q_R\rangle=g(a_x^2-a_y^2)/(4\omega_R^{(0)2})$, while the frequency of degenerate IR mode will be split into $\sqrt{\omega_{IR}^{(0)2}\mp g\langle Q_R\rangle}$, which is similar to Eq.(\ref{Eq_Sol_shift_freq_2}) in that the last two terms are neglected. This frequency splitting then leads to nonzero phonon angular momentum $L_{IR,z}$ and magnetization $M_z=\mu_{ph}L_{IR,z}/(\hbar V_c)$, which oscillate in superposition state of frequencies $(\omega_{IR,y}\mp\omega_{IR,x})/2$ with ampltiudes $a_xa_y(\omega_{IR,y}\pm\omega_{IR,x})/2$, as shown in Fig.\ref{Fig_QIRx2_y2}(a2,b2,c2). Here $\mu_{ph}$ is the phonon magnetic moment, $\hbar$ is the reduced Planck constant, and $V_c$ is the unit cell volume. If the Raman rectification leads to ferroelectric phase and the phonon magnetic moment is of order $\mu_B$, the dynamical multiferroicity will occur.

Furthermore, it is also possible to induce Raman rectification and oscillating magnetization simultaneously by linearly polarized mid-IR pulse not along $x$,  $y$ or $x\pm y$ direction, as shown in Fig.\ref{Fig_QIRx2_y2}(a4,b4,c4).  The sign of Raman rectification can be switched by changing the long axis of elliptically polarized pulse to along $y$, or the angle of linearly polarized pulse, while the direction of angular momentum  in the first half-cycle can be switched by changing the handedness of elliptically polarized pulse, or using the $x$/$y$-symmetric partner of linearly polarized pulse. 

For circularly polarized pulse ($\phi=\pi/4$), $Q_{IR,x}$ and $Q_{IR,y}$ are excited to the same amplitudes ($a_x=a_y$), and the rectification of Raman mode disappears (see Table.\ref{tab:Frequency_Amplitude_2}). The frequency of IR mode does not split, but increases slightly as the strength of pump pulse increases. The IR mode has angular momentum with amplitude $a_x^2 \omega_{IR,x}$. 

\subsection{Discussion}
\label{Discussion}
We note that the saturation of Raman rectification are more easily observed in systems with larger coupling constant. This can be seen as follows, suppose $Q_{IR}$ and $Q_{R}$ are solutions of Eq.(\ref{Eq_EOM}) at coupling constant $g$ and pump field $E_{0}$, if $g\rightarrow kg$, where $k$ is a constant, then $Q_{IR}\rightarrow \frac{1}{k}Q_{IR}$, $Q_{R}\rightarrow \frac{1}{k}Q_{R}$ and $E_{0}\rightarrow \frac{1}{k}E_{0}$ will also be a solution for the new coupling constant. This means that both the saturation amplitude (if it exists for the two-phonon system) and the required pump field to realize saturation will be reduced.

The frequency renormalization of IR mode can still be observed if damping effect is included. The corresponding Fourier transform of $Q_{IR}$ will be asymmetrically peaked at $\omega_{IR}$ and broadens as $E_0$ increases. The rectification behavior of Raman mode is somehow difficult to be discerned --- the peak of $Q_R$ near $t=0$ increases as $E_0$ increases, but the peak of Fourier transform of $Q_R$ at $0$ tends to saturate and broadens at large $E_0$. It should be pointed out that the peak of $Q_R$ near $t=0$ is not the rectified Raman mode we've discussed above. As can be seen in Fig.\ref{Fig_QR_QIR_t_without_damping}(c), the Raman mode also has a peak near $t=0$ at large $E_0$, but that's not the rectified Raman mode, which is the rectified component of long-time solution $Q_R(t)$, where $E(t)$ approaches zero.

Finally, the discussion on ``dynamically induced multiferroicity'' can be applied to systems with coupling of form $Q'_{IR,x}Q'_{IR,y}Q_R$, simply by coordinate transformation $Q_{IR,x}=(Q'_{IR,x}+Q'_{IR,y})/\sqrt{2}$ and $Q_{IR,y}=(Q'_{IR,x}-Q'_{IR,y})/\sqrt{2}$. The ``overshoot of magnetization upon decay'' in Paiva \textit{et. al.}'s work\cite{7lm1-wm3y} reveals the hidden oscillating behavior of magnetization discussed above, covered up by damping effect.

\section{Conclusion}
In conclusion, we have given an analytical solution of the two-phonon system coupled by $gQ_RQ_{IR}^2$ to second order of coupling constant $g$. The frequency renormalization of IR mode suppresses the linear increase of $a_{IR}$ with driving amplitude $E_0$, eventually leading to a saturation of the rectified amplitude of the Raman phonon. This finding suggests that under the mechanism of anharmonic coupling, it might not be possible to drive a crystalline system to any displacement along the eigenvector of Raman phonon before the breaking of the system, even though it is allowed by the potential energy landscape. For degenerate IR mode, the frequency splitting leads to simultaneous Raman rectification and oscillating magnetization, when pumped by elliptically or linearly polarized pulse. The saturation phenomena can be tested on systems with large anharmonic coupling constant by femtosecond x-ray measurement. Our work extends the dynamical multiferroicity from systems with coupling $Q_{IR,x}Q_{IR,y}Q_R$ to systems with coupling $Q_R(Q_{IR,x}^2-Q_{IR,y}^2)$, and the method can be readily applied to other anharmonic terms, \textit{e.g.} $Q_R^2Q_{IR}^2$.

\begin{acknowledgments}
 Y.Z. would like to thank H. Mao for discussions about Fast Fourier transformation and the National Supercomputer Center in Guangzhou for computational supports. This work is supported by Zhejiang Provincial Natural Science Foundation (Grant No. LQ21A040010).
\end{acknowledgments}

\appendix
\section{Deduction of frequency-amplitude relations}
\label{Apend1}
Suppose the solution of Eq. (\ref{Eq_EOM}) can be written as a series of successive approximations:
\begin{subequations}
	\label{Eq_Q}
	\begin{equation}
	Q_{R}=Q_{R}^{(0)}+Q_{R}^{(1)}+Q_{R}^{(2)}+\cdots
	\end{equation}
	\begin{equation}
	Q_{IR}=Q_{IR}^{(0)}+Q_{IR}^{(1)}+Q_{IR}^{(2)}+\cdots
	\end{equation}
\end{subequations}
where $Q_{R}^{(0)}=a_Rcos(\omega_{R}t)$ and $Q_{IR}^{(0)}=a_{IR}cos(\omega_{IR}t)$. The difference in initial phases between $Q_R^{(0)}$ and $Q_{IR}^{(0)}$ does not influence the following results and is taken to be zero. The superscript $(i)$ denotes that this term is proportional to  the $i$-th power of the coupling constant $g$. The frequencies $\omega_{R}$ and $\omega_{IR}$ are exact and can be expanded similarly as 
\begin{subequations}
	\label{Eq_omega}
	\begin{equation}
	\omega_{R}=\omega_{R}^{(0)}+\omega_{R}^{(1)}+\omega_{R}^{(2)}+\cdots
	\end{equation}
	\begin{equation}
	\omega_{IR}=\omega_{IR}^{(0)}+\omega_{IR}^{(1)}+\omega_{IR}^{(2)}+\cdots
	\end{equation}
\end{subequations}
Substitute Eq.(\ref{Eq_Q}) and Eq.(\ref{Eq_omega}) into Eq.(\ref{Eq_EOM}) and equate the coefficients of identical powers of $g$, recursively solvable equations will be obtained. Retaining terms of the first-order,  we obtain for $Q_R^{(1)}$ and $Q_{IR}^{(1)}$ the equations
\begin{subequations}
	\label{Eq_EOM_1st}
	\begin{equation}
	\ddot{Q}_R^{(1)}+\omega_{R}^{(0)2}Q_R^{(1)}=\frac{1}{2}ga_{IR}^2cos^2(\omega_{IR}t)+2\omega_{R}^{(0)}\omega_{R}^{(1)}a_Rcos(\omega_{R}t)
	\end{equation}
	\begin{equation}
	\begin{aligned}
	\ddot{Q}_{IR}^{(1)}+\omega_{IR}^{(0)2}Q_{IR}^{(1)}=ga_Ra_{IR}cos(\omega_{R}t)cos(\omega_{IR}t)\\+2\omega_{IR}^{(0)}\omega_{IR}^{(1)}a_{IR}cos(\omega_{IR}t)
	\end{aligned}
	\end{equation}
\end{subequations}
By requiring the resonant terms ($cos(\omega_{R}t)$ and $cos(\omega_{IR}t)$) should vanish in Eq.(\ref{Eq_EOM_1st}), we have $\omega_{R}^{(1)}=\omega_{IR}^{(1)}=0$, and
\begin{subequations}
	\label{Eq_Sol_1st}
	\begin{equation}
		Q_R^{(1)}=\frac{\frac{1}{4}ga_{IR}^2}{\omega_{R}^{(0)2}}+\frac{\frac{1}{4}ga_{IR}^2}{\omega_{R}^{(0)2}-(2\omega_{IR}^{(0)})^{2}}cos(2\omega_{IR}t)
	\end{equation}
	\begin{equation}
		Q_{IR}^{(1)}=\frac{1}{2}ga_Ra_{IR}[\frac{cos(\omega_{IR}-\omega_{R})t}{\omega_{IR}^{(0)2}-(\omega_{IR}^{(0)}-\omega_{R}^{(0)})^2}+\frac{cos(\omega_{IR}+\omega_{R})t}{\omega_{IR}^{(0)2}-(\omega_{IR}^{(0)}+\omega_{R}^{(0)})^2}]
	\end{equation}
\end{subequations}
It is obvious from Eq.(\ref{Eq_Sol_1st}a) that $Q_R$ will oscillate in $2\omega_{IR}$ around a displaced position, or $Q_R$ is rectified, even if $Q_R$ is initially at rest ($(Q_R,\dot{Q}_R)=(0,0)$). To obtain better rectification, the amplitude of displacement should be much larger than those of oscillations, \textit{i.e.} $\sqrt{2}\omega_{IR}^{(0)}\gg \omega_{R}^{(0)}$ and $\frac{\frac{1}{4}ga_{IR}^2}{\omega_{R}^{(0)2}}\gg a_R$. Eq.(\ref{Eq_Sol_1st}b) indicates that additional oscillations with frequencies $\omega_{IR}\pm\omega_{R}$ are superposed on eigen mode of  $Q_{IR}$\cite{Ionic_Raman_1974}.

In a similar manner, by retaining terms of the second-order, we obtain equations for $Q_{R}^{(2)}$ and $Q_{IR}^{(2)}$,
\begin{subequations}
	\label{Eq_EOM_2nd}
	\begin{equation}
		\ddot{Q}_R^{(2)}+\omega_{R}^{(0)2}Q_{R}^{(2)}=ga_{IR}cos(\omega_{IR}t)Q_{IR}^{(1)}+2\omega_{R}^{(0)}\omega_{R}^{(2)}a_Rcos(\omega_{R}t)
	\end{equation}
	\begin{equation}
	\begin{aligned}
	\ddot{Q}_{IR}^{(2)}+\omega_{IR}^{(0)2}Q_{IR}^{(2)}&=g(a_Rcos(\omega_{R}t)Q_{IR}^{(1)}+a_{IR}cos(\omega_{R}t)Q_{R}^{(1)})\\
	&+2\omega_{IR}^{(0)}\omega_{IR}^{(2)}a_{IR}cos(\omega_{IR}t)
	\end{aligned}
	\end{equation}
\end{subequations}
Substitute Eq.(\ref{Eq_Sol_1st}) into the right side of Eq.(\ref{Eq_EOM_2nd}) and let the resonant terms vanish, we then obtain the shift of frequencies as a function of the amplitudes as shown in the main text, Eq.(\ref{Eq_Sol_shift_freq}). The remaining terms on the right side of Eq.(\ref{Eq_EOM_2nd}) then give
\begin{equation}
\label{Eq_Sol_2nd_a}
\begin{aligned}
Q_{R}^{(2)}&=\frac{1}{4}g^2a_Ra_{IR}^2\{\frac{cos[(2\omega_{IR}-\omega_{R})t]}{[\omega_{IR}^{(0)2}-(\omega_{IR}^{(0)}-\omega_{R}^{(0)})^2][\omega_{R}^{(0)2}-(2\omega_{IR}^{(0)}-\omega_{R}^{(0)})^2]}\\
&+\frac{cos[(2\omega_{IR}+\omega_{R})t]}{[\omega_{IR}^{(0)2}-(\omega_{IR}^{(0)}+\omega_{R}^{(0)})^2][\omega_{R}^{(0)2}-(2\omega_{IR}^{(0)}+\omega_{R}^{(0)})^2]}\}
\end{aligned}
\end{equation}
and 
\begin{equation}
\label{Eq_Sol_2nd_b}
\begin{aligned}
Q_{IR}^{(2)}&=\frac{1}{4}g^2a_R^2a_{IR}\{\frac{cos[(\omega_{IR}-2\omega_{R})t]}{[\omega_{IR}^{(0)2}-(\omega_{IR}^{(0)}-\omega_{R}^{(0)})^2][\omega_{IR}^{(0)2}-(\omega_{IR}^{(0)}-2\omega_{R}^{(0)})^2]}\\
&+\frac{cos[(\omega_{IR}+2\omega_{R})t]}{[\omega_{IR}^{(0)2}-(\omega_{IR}^{(0)}+\omega_{R}^{(0)})^2][\omega_{IR}^{(0)2}-(\omega_{IR}^{(0)}+2\omega_{R}^{(0)})^2]}\}\\
&+\frac{1}{8}g^2a_{IR}^3\frac{cos(3\omega_{IR})t}{[\omega_{R}^{(0)2}-(2\omega_{IR}^{(0)})^2][\omega_{IR}^{(0)2}-(3\omega_{IR}^{(0)})^2]}
\end{aligned}
\end{equation}
Eqs.(\ref{Eq_Sol_2nd_a}) and (\ref{Eq_Sol_2nd_b}) show that up to second order of smallness, additional components with frequencies $2\omega_{IR}\pm\omega_{R}$ are added on $Q_R$, and components with frequencies $\omega_{IR}\pm2\omega_{R}$ and $3\omega_{IR}$ are superposed on $Q_{IR}$.

The above processes can be extended to higher orders and more modes will occur. For example, $\omega_{R}^{(3)}=\omega_{IR}^{(3)}=0$, $Q_R^{(3)}$ contains additional oscillations of frequencies $2\omega_{IR}\pm2\omega_{R}$, $2\omega_{R}$, and $4\omega_{IR}$, and $Q_{IR}^{(3)}$ contains additional frequencies $\omega_{IR}\pm3\omega_{R}$ and $3\omega_{IR}\pm\omega_{R}$. However, their amplitudes are much weaker and will not be considered here.

\section{The case of ionic Raman scattering}
\label{Append_ionicRaman}
\begin{figure}[h]
		\includegraphics[width=8cm]{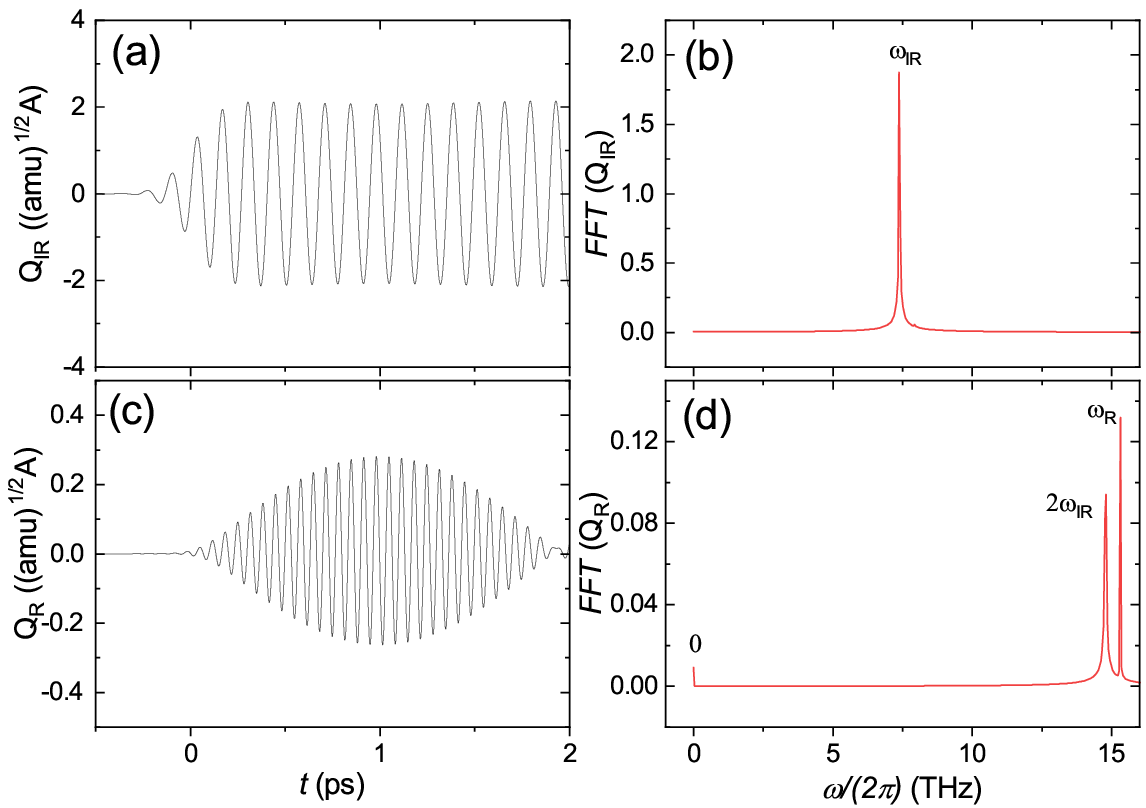}
		\caption{ (a) $Q_{IR}(t)$ and (c) $Q_R(t)$ under pump pulse with $E_0=8$ MV/cm, $\omega_{IR}^{(0)}/(2\pi)=7.4$ THz, $\omega_{R}^{(0)}/(2\pi)=15.3$ THz, other parameters are the same with Juraschek \textit{et. al.}'s work on BiFeO$_3$\cite{PhysRevB.97.174302}. (b) and (d) are the amplitudes of Fourier transform of long-time solutions of $Q_{IR}$ and $Q_R$, respectively.}
		\centering
		\label{Fig_QR_QIR_t_SumFreq}
\end{figure}

Consider the special case when $\omega_{IR}^{(0)}\approx\frac{1}{2}\omega_{R}^{(0)}$, which we show corresponds to the ``sum-frequency ionic Raman scattering"\cite{PhysRevB.97.174302}. By referring to Table.\ref{tab:Frequency_Amplitude} it's clear that $Q_R^{(1)}(2\omega_{IR})$ has large amplitude due to the denominator $\omega_{R}^{(0)2}-(2\omega_{IR}^{(0)})^2$ approaches zero. Meanwhile, the $Q_R^{(0)}(\omega_R)$ mode will also be excited, with $a_R$ of the same magnitude with $\frac{1}{4}ga_{IR}^2/(\omega_{R}^{(0)2}-(2\omega_{IR}^{(0)})^2)$. As a result, $Q_R$ will oscillate in slightly different frequencies ($\omega_{R}$ and $2\omega_{IR}$) and form beats, with beat frequency $\omega_{R}-2\omega_{IR}$, see Fig.\ref{Fig_QR_QIR_t_SumFreq} for the coupled $A_1$ modes ($15.3$ THz for Raman and $7.4$ THz for IR phonons) of BiFeO$_3$\cite{PhysRevB.97.174302}. Lastly, the rectification of Raman mode can be neglected since the rectification amplitude $\frac{1}{4}ga_{IR}^2/\omega_{R}^{(0)2}$ is much smaller than that of $2\omega_{IR}$ mode $\frac{1}{4}ga_{IR}^2/(\omega_{R}^{(0)2}-(2\omega_{IR}^{(0)})^2)$. 

\section{The case of degenerate IR mode}
\label{Apend3}
\begin{table}[H]
	\caption{The angular frequencies and amplitudes of components for $Q_R$, $Q_{IR,x}$ and $Q_{IR,y}$ to the first-order. The expressions for $\omega_{IR,x}$ and $\omega_{IR,y}$ are given by Eqs.(\ref{Eq_Sol_shift_freq_2}).}
	\begin{ruledtabular}
		\begin{tabular}{ccc}
			Term & Angular frequency  & Amplitude  \\
			\hline
			$Q_R^{(0)}$ & $\omega_{R}$ & $a_R$  \\
			$Q_R^{(1)}$ & $0$ & $\frac{\frac{1}{4}g(a_{x}^2-a_{y}^2)}{\omega_{R}^{(0)2}}$  \\
			 & $2\omega_{IR,x}$ & $\frac{\frac{1}{4}ga_{x}^2}{\omega_{R}^{(0)2}-(2\omega_{IR}^{(0)})^2}$  \\
			 & $2\omega_{IR,y}$ & $\frac{\frac{1}{4}ga_{y}^2}{\omega_{R}^{(0)2}-(2\omega_{IR}^{(0)})^2}$  \\
			 \dots & \dots & \dots \\
			  \hline
			  $Q_{IR,x}^{(0)}$ & $\omega_{IR,x}$ & $a_{x}$  \\
			  $Q_{IR,x}^{(1)}$ & $\omega_{IR,x}-\omega_{R}$ & $\frac{\frac{1}{2}ga_Ra_{x}}{\omega_{IR}^{(0)2}-(\omega_{IR}^{(0)}-\omega_{R}^{(0)})^2}$  \\
			   & $\omega_{IR,x}+\omega_{R}$ & $\frac{\frac{1}{2}ga_Ra_{x}}{\omega_{IR}^{(0)2}-(\omega_{IR}^{(0)}+\omega_{R}^{(0)})^2}$  \\
			   \dots & \dots & \dots \\
			    \hline
			  $Q_{IR,y}^{(0)}$ & $\omega_{IR,y}$ & $a_{y}$  \\
			  $Q_{IR,y}^{(1)}$ & $\omega_{IR,y}-\omega_{R}$ & $\frac{\frac{1}{2}ga_Ra_{y}}{\omega_{IR}^{(0)2}-(\omega_{IR}^{(0)}-\omega_{R}^{(0)})^2}$  \\
			   & $\omega_{IR,y}+\omega_{R}$ & $\frac{\frac{1}{2}ga_Ra_{y}}{\omega_{IR}^{(0)2}-(\omega_{IR}^{(0)}+\omega_{R}^{(0)})^2}$  \\
			     \dots & \dots & \dots \\
		\end{tabular}
	\end{ruledtabular}
	\label{tab:Frequency_Amplitude_2}
\end{table}

\section{The quantum mechanical treatment}
\label{Apend2}
In this section we show that the frequency renormalization can be obtained by perturbation method to the second order. The Hamiltonian of the coupled two-phonon system is
\begin{equation}
        \label{Eq_H}
	\hat{H}=\hat{H}_{0}+\hat{H}_{1}
\end{equation}
where $\hat{H}_{0}=\frac{1}{2}\hat{P}_{R}^2+\frac{1}{2}\omega_{R}^{(0)2}\hat{Q}_{R}^2+\frac{1}{2}\hat{P}_{IR}^2+\frac{1}{2}\omega_{IR}^{(0)2}\hat{Q}_{IR}^2$ and $\hat{H}_{1}=-\frac{1}{2}g\hat{Q}_{R}\hat{Q}_{IR}^2$. Here $\hat{Q}$ has the dimension of mass$^{1/2}\times$length, and $\hat{H}_0$ corresponds to a two-dimensional harmonic oscillator with eigenvalues $E_{n,m}^{(0)}=(n+\frac{1}{2})\hbar\omega_{R}^{(0)}+(m+\frac{1}{2})\hbar\omega_{IR}^{(0)}, n,m=0,1,2,\dots$. The eigenfunctions are $\psi_{n,m}(Q_R,Q_{IR})=u_n(Q_R)u_m(Q_{IR})$, with
\begin{equation}
	u_{n}(Q_i)=N_nH_n(\alpha_iQ_i)e^{-\frac{1}{2}\alpha_i^2Q_i^2}
\end{equation}
where $N_n=(\frac{\alpha_i}{\sqrt{\pi}2^nn!})^{1/2}$ is the normalization constant, $\alpha_i=(\omega_{i}^{(0)}/\hbar)^{1/2}, i=R,IR$ and $H_n$ is the $n$th Hermite polynominal.

The eigenvalues of $\hat{H}$ up to second-order are
\begin{equation}
\label{Energy_append}
	E_{n,m}=E_{n,m}^{(0)}+\bra{n,m}\hat{H}_1\ket{n,m}+\sum_{n',m'\neq n,m}\frac{|\bra{n,m}\hat{H}_1\ket{n',m'}|^2}{E_{n,m}^{(0)}-E_{n',m'}^{(0)}}
\end{equation}
By virtue of 
\begin{equation}
	\bra{n}\hat{Q}_i\ket{m}=\left\{
	\begin{aligned}
	&\frac{1}{\alpha_i}\sqrt{\frac{n}{2}} ,& m=n-1 \\
	&\frac{1}{\alpha_i}\sqrt{\frac{n+1}{2}} ,& m=n+1 \\
	&0 ,& otherwise
	\end{aligned}
	\right. 
\end{equation}
and
\begin{equation}
\bra{n}\hat{Q}_i^2\ket{m}=\left\{
\begin{aligned}
&\frac{1}{2\alpha_i^2}\sqrt{(n-1)n} ,& m=n-2 \\
&\frac{1}{2\alpha_i^2}\sqrt{2n+1} ,& m=n \\
&\frac{1}{2\alpha_i^2}\sqrt{(n+1)(n+2)} ,& m=n+2 \\
&0 ,& otherwise
\end{aligned}
\right. 
\end{equation}
the first-order term in Eq.(\ref{Energy_append}) vanishes. There are only six second-order terms, corresponding to $n'=n\pm 1$ and $m'=m,m\pm 2$. A little algebraic efforts lead to 
\begin{equation}
E_{n,m}=\hbar\omega_{R}(n+\frac{1}{2})+\hbar\omega_{IR}(m+\frac{1}{2})+\Delta E_0
\end{equation}
where
\begin{widetext}
\begin{subequations}
	\begin{equation}
	\omega_{R}=\omega_{R}^{(0)}-\frac{g^2\bra{m}Q_{IR}^2\ket{m}}{4\omega_{R}^{(0)}}\frac{1}{(2\omega_{IR}^{(0)})^2-(\omega_{R}^{(0)})^2}
	\end{equation}
	\begin{equation}
	\omega_{IR}=\omega_{IR}^{(0)}-\frac{g^2\bra{m}Q_{IR}^2\ket{m}}{8\omega_{IR}^{(0)}}[\frac{1}{\omega_{R}^{(0)2}}-\frac{1}{2}\frac{1}{(2\omega_{IR}^{(0)})^2-\omega_{R}^{(0)2}}]-\frac{g^2\bra{n}Q_{R}^2\ket{n}}{4\omega_{IR}^{(0)}}\frac{1}{(2\omega_{IR}^{(0)})^2-(\omega_{R}^{(0)})^2}
	\end{equation}
	\begin{equation}
		\Delta E_0=\frac{3}{64}\frac{g^2\hbar^2}{\omega_{IR}^{(0)2}}\frac{1}{(2\omega_{IR}^{(0)})^2-(\omega_{R}^{(0)})^2}
	\end{equation}
\end{subequations}
\end{widetext}
It's clear that the frequency renormalization has the same form with Eq.(\ref{Eq_Sol_shift_freq}), if $a_{IR}$ is replaced by $|\bra{m}Q_{IR}^2\ket{m}|^{1/2}$ and $a_R$ by $|\bra{n}Q_R^2\ket{n}|^{1/2}$. Different from the classical treatment, the initial values of $Q_R$ and $Q_{IR}$ can not be assumed zero due to the zero-point energy, and can be estimated by requiring that $\frac{1}{2}\omega_{i}^{(0)2}Q_{i}^2=\frac{1}{2}\hbar\omega_{i}^{(0)}$, which then gives $Q_{i}|_{t_0}=\sqrt{\hbar/\omega_{i}^{(0)}}, i=R, IR$.

\bibliography{Nonlinear_QRQIR2_bib}

\end{document}